\documentclass[sigconf, nonacm]{acmart} 

\usepackage[frozencache,cachedir=.]{minted}
\usepackage{multicol}
\usepackage{enumitem}
\usepackage{todonotes}
\usepackage{rotating}
\usepackage{multirow}
\usepackage{natbib}
\usepackage{wasysym}
\RequirePackage[skins]{tcolorbox}
\newtcolorbox{custombox}[1]{
	colback=white,
	colframe=white,
	left=1mm,
	right=1mm,
	top=1mm,
	bottom=1mm,
	fonttitle=\bfseries,
	arc=0mm,
	leftrule=1mm,
	rightrule=0mm,
	toprule=0mm,
	bottomrule=0mm,
	notitle,
	before=\par\smallskip\noindent,
	before upper={\textbf{#1: } },
}

\AtBeginDocument{%
  \providecommand\BibTeX{{%
    \normalfont B\kern-0.5em{\scshape i\kern-0.25em b}\kern-0.8em\TeX}}}

\copyrightyear{2024} 
\acmYear{2024} 
\setcopyright{rightsretained}
\acmConference[SIGCSE 2024] {Proceedings of the 55th ACM Technical Symposium on Computer Science Education V. 1}{March 20--23, 2024}{Portland, OR, United States.} 
\acmBooktitle{Proceedings of the 55th ACM Technical Symposium on Computer Science Education V. 1 (SIGCSE 2024), March 20--23, 2023, Portland, OR, United States} 
\acmPrice{15.00}
\acmDOI{XXXXXXX.XXXXXXX}
\acmISBN{XXX-X-XXXX-XXXX-X/XX/XX}

\begin{document}
\pagestyle{plain} 



\title{Promptly: Using Prompt Problems to Teach Learners How to Effectively Utilize AI Code Generators}


\author{Paul Denny}
\orcid{0000-0002-5150-9806}
\affiliation{
  \institution{The University of Auckland}
  \city{Auckland}
  \country{New Zealand}
}
\email{paul@cs.auckland.ac.nz}

\author{Juho Leinonen}
\orcid{0000-0001-6829-9449}
\affiliation{
  \institution{The University of Auckland}
  \city{Auckland}
  \country{New Zealand}
}
\email{juho.leinonen@auckland.ac.nz}

\author{James Prather}
\orcid{0000-0003-2807-6042}
\affiliation{
  \institution{Abilene Christian University}
  \city{Abilene, TX}
  \country{USA}
}
\email{james.prather@acu.edu}

\author{Andrew Luxton-Reilly}
\orcid{0000-0001-8269-2909}
\affiliation{
  \institution{The University of Auckland}
  \city{Auckland}
  \country{New Zealand}
}
\email{a.luxton-reilly@auckland.ac.nz}

\author{Thezyrie Amarouche}
\orcid{0000-0003-3725-0049}
\affiliation{
  \institution{University of Toronto Scarborough}
  \city{Toronto, ON}
  \country{Canada}
}
\email{thezyrie.amarouche@mail.utoronto.ca}

\author{Brett A. Becker}
\orcid{0000-0003-1446-647X}
\affiliation{
  \institution{University College Dublin}
  \city{Dublin}
  \country{Ireland}
}
\email{brett.becker@ucd.ie}

\author{Brent N. Reeves}
\orcid{0000-0001-5781-1136}
\affiliation{%
  \institution{Abilene Christian University}
  \city{Abilene, TX}
  \country{USA}
}
\email{brent.reeves@acu.edu}

\renewcommand{\shortauthors}{Authors, et al.}

\begin{abstract}
With their remarkable ability to generate code, large language models (LLMs) are a transformative technology for computing education practice.  They have created an urgent need for educators to rethink pedagogical approaches and teaching strategies for newly emerging skill sets.
Traditional approaches to learning programming have focused on frequent and repeated practice at \emph{writing} code.  The ease with which code can now be generated has resulted in a shift in focus towards reading, understanding and evaluating LLM-generated code.  
In parallel with this shift, a new essential skill is emerging -- the ability to construct good prompts for code-generating models.
This paper introduces a novel pedagogical concept known as a `Prompt Problem', designed to help students learn how to craft effective prompts for LLMs. A Prompt Problem challenges a student to create a natural language prompt that leads an LLM to produce the correct code for a specific problem.  To support the delivery of Prompt Problems at scale, in this paper we also present a novel tool called \textsc{Promptly} which hosts a repository of Prompt Problems and automates the evaluation of prompt-generated code.
We report empirical findings from a field study in which Promptly was deployed in a first-year Python programming course ($n=54$).  We explore student interactions with the tool and their perceptions of the Prompt Problem concept.
We found that Promptly was largely well-received by students for its ability to engage their computational thinking skills and expose them to new programming constructs. 
We also discuss avenues for future work, including variations on the design of Prompt Problems and the need to study their integration into the curriculum and teaching practice. 

\end{abstract}

\maketitle

%
%
\section{Introduction}


The advent of large language models (LLMs) that can generate code is having a rapid and significant impact on computing education practice, particularly at the introductory level.  Traditional pedagogical approaches have focused on helping students learn how to \emph{write} code.  This is typically achieved through frequent practice involving many small problems \cite{allen2019many, denny2011codewrite} or through scaffolding via activities such as Parsons problems \cite{ericson2022parsons, du2020review}.  However, LLMs are now capable of producing code automatically and have demonstrated impressive performance on problems that are typical in introductory programming courses \cite{finnie-ansley2022robots, finnie-ansley2023my, reeves2023evaluating}. 
In addition to the opportunities they present, educators have voiced concerns around the potential misuse of these models for plagiarism, and over-reliance on AI-generated code by beginners~\cite{becker2023programming}, leading to a possible erosion of traditional coding skills \cite{denny2023computing}. New pedagogical approaches are needed to develop the changing skillsets that students require in the era of generative AI \cite{denny2023chat}. 


Teaching students to read and understand code are longstanding goals of introductory courses, and they are becoming increasingly important skills given the ease with which code can be generated by LLM-based tools. 
An equally important emerging skill is the ability to formulate effective prompts for LLMs to generate code. Recent work has shown that although many typical introductory problems can be solved by LLMs using verbatim textbook or exam problem statements~\cite{finnie-ansley2022robots, finnie-ansley2023my}, this approach is not always sufficient.  For example, manual modification of the prompts to include explicit algorithmic hints greatly improves code-generation performance \cite{tang2022solving}.  In recent work, Denny et al. argue that the ability to engineer effective prompts that generate correct solutions is now an essential skill for students, yet they do not propose concrete approaches for how this skill can be taught \cite{denny2023conversing}.

To address this concern, in the current paper we introduce the concept of a `Prompt Problem' -- a new exercise paradigm in which students solve programming exercises by formulating natural language prompts for code-generating LLMs.  Students are presented with a visual representation of a problem that illustrates how input values should be transformed to an output.  Their task is to devise a prompt that would guide an LLM to generate the code required to solve the problem.  

In addition to conceptualizing the problem type, we make two other contributions: we introduce a tool for delivering Prompt Problems and we present empirical findings from the use of this tool by introductory progamming students.  To understand how Prompt Problems work in practice, we have developed a web-based tool called \textsc{Promptly}.  This tool displays a problem representation, converts a prompt written by a student to code (via an API call to an LLM), and then executes the code against a suite of test cases.  If the code fails to solve the problem, the student must revise and resubmit their prompt.  This iterative process continues until the problem is solved.  We hypothesize that solving Prompt Problems will help students learn how to 
craft effective prompts.

We begin by presenting an illustrative example of a Prompt Problem, and we draw from the results of a pilot study to demonstrate the difficulty that students commonly face in formulating effective prompts.  We then describe the design of our tool, \textsc{Promptly}, for delivering Prompt Problems at scale and we deploy it in an introductory Python programming course ($n=54$).  We report the results of our analysis of student interactions with the tool and their perceptions of the activity.  We also discuss possible variations of the problem design, and suggest ideas for future work.

\section{Related work}
It has been less than a year since LLMs began to dominate conversations in the computing education community and a little more than that since the first research papers began to emerge in the computing education literature. Early work centered on the capabilities of these tools, largely driven by concerns that they would lead to a flood of cheating~\cite{malinka2023on} and the effect that would have on student learning. Sometimes, such work involved comparing LLM and student performance, for example in generating explanations of code~\cite{leinonen2023comparing}. Finnie-Ansley et al. demonstrated that Codex (based on GPT-3) ranked in the top quartile of real introductory programming (CS1) students on real exams~\cite{finnie-ansley2022robots}. A year later Finnie-Ansley et al. extended this work to data structures and algorithms (CS2) exams with very similar results~\cite{finnie-ansley2023my}. Other studies on the capabilities of LLMs have revealed impressive proficiency in dealing with object-oriented programming tasks~\cite{cipriano2023gpt-3}, Parsons problems~\cite{reeves2023evaluating}, and compiler error messages~\cite{leinonen2023using}. Many of these explorations also revealed that LLMs are not infallible and can produce solutions that don't align with best programming practice~\cite{cipriano2023gpt-3}, struggle with longer and higher-level specifications~\cite{finnie-ansley2022robots}, include unnecessary elements~\cite{wermelinger2023using}, and cause students to become confused reading code that they didn't write themselves~\cite{kazemitabaar2023studying, prather2023its}. Babe et al. showed that LLMs can mislead students, causing them to believe that their own prompts are more (or less) effective than they are in reality~\cite{babe2023studenteval}.

Recently, the focus has started to shift from assessing the capabilities of LLMs to using them in teaching and learning practice~\cite{macneil2023implications, moore2023empowering}. Sarsa et al. showed that LLMs can generate viable programming questions including test cases and explanations~\cite{sarsa2022automatic}. Complementing this reshaping of the practices of teaching and learning, the importance of details such as context~\cite{leinonen2023using} and prompting~\cite{denny2023conversing} have begun to emerge. For example, White et al. present a prompt structuring framework for constructing prompts so they can be applied across problem domains, a catalog of prompts that have been successfully applied to improve LLM responses, and a demonstration of how prompts can be constructed from patterns and how such patterns can be combined effectively \cite{white2023prompt}. There is increasing interest in understanding the types of prompts that students construct when communicating with LLMs.  Babe et al. developed a benchmark dataset of 1,749 prompts aimed at 48 problems, written by 80 novice Python programming students~\cite{babe2023studenteval} which can be used by others for LLM benchmarking as well as tool development.

A logical next step towards integrating LLMs into teaching practice is developing tools and resources to aid students in effectively working with LLMs for learning. Lao and Guo interviewed 19 introductory programming instructors from nine countries across six continents and found that some instructors are embracing the idea of integrating AI tools into current courses via mechanisms such as giving personalized help to students and aiding instructors with time-consuming tasks~\cite{lau2023ban}. MacNeil et al. used LLM-generated code explanations successfully in a web software development e-book~\cite{macneil2023experiences}, and Zingaro and Porter are completing a textbook for teaching introductory programming with Copilot and ChatGPT from day one~\cite{Porter2023ai}. Integrating LLMs into computer science courses seems inevitable and stands to transform the way the subject is taught at all levels \cite{tedre2023k12computing,denny2023chat}. We believe that Prompt Problems will be one important step along the journey towards integrating the use of LLMs in computer science education.

\begin{figure}
\centering
  \includegraphics[width=.8\linewidth]{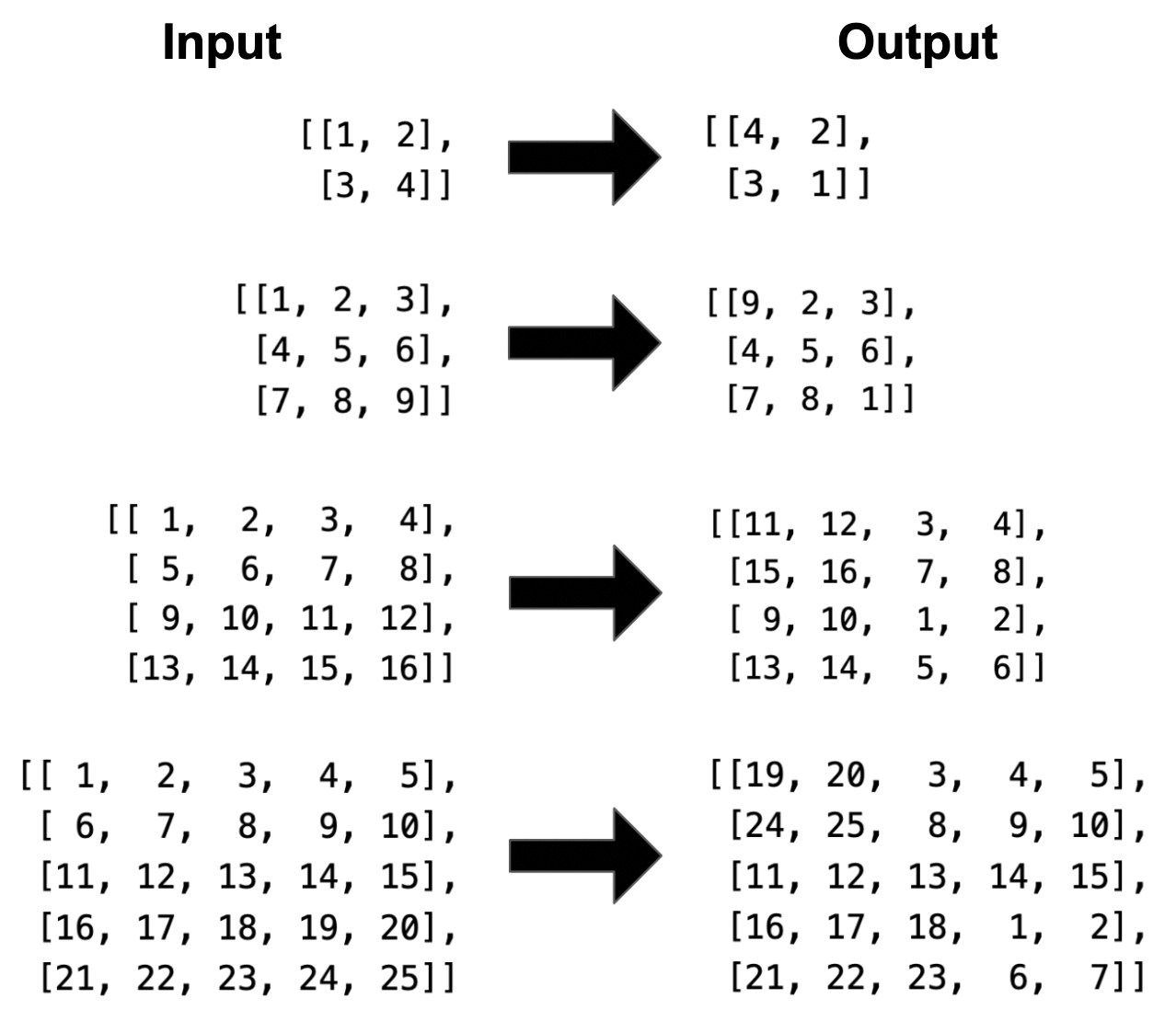}
  \caption{An example Prompt Problem that displays the data visually so that students cannot simply copy and paste the description into an LLM.  The goal is to swap the top-left and bottom-right non-overlapping quadrants of the matrix.}
  \label{fig:array-swap-problem}
\end{figure}

%
%
\section{Pilot Study}
\label{sec:pilot_study}

In order to understand how students might try to use LLM tools like ChatGPT to communicate program requirements, we asked a group of graduate students at the University of Auckland to participate in a prompt writing assignment. This assignment took place during a single class session in April 2023. We provided a visual representation of a problem (see Figure \ref{fig:array-swap-problem}) and asked participants to query ChatGPT to write a program that could convert the shown input to the shown output. The problem description was provided visually to prevent participants from easily copying and pasting it and, instead, to encourage them to formulate a suitable prompt themselves. We also asked students to keep a log of their interactions with ChatGPT, organized by their prompt followed by ChatGPT's response, etc. in sequential order. We hoped that this would provide information about the challenges that students face when using LLMs to write code, and to suggest where a tool might potentially help to address those difficulties. 

Thirty-six graduate students participated in the pilot study by completing the activity described above and reflecting on it by writing an open-response review of the task. Fifteen participants chose to provide us with their ChatGPT conversations. We discuss these below in Section \ref{sec:pilot-interaction}. Thirty-one participants shared their reflections on the task, which we discuss in Section \ref{sec:pilot-perceptions}.

\subsection{Interactions}
\label{sec:pilot-interaction}

%
%

Listing \ref{listing:chat_1} shows an example of a student in the pilot study going back-and-forth with ChatGPT when trying to solve the problem.  Notice the `conversational' beginning and the subsequent iterative refinement.  Each successive prompt attempts to refine the code that is already generated, rather than modify the original prompt to be more complete.
We expected computer science graduate students to have very few problems writing effective prompts, however several interesting trends emerged from this activity. 
First, many students began with an incomplete prompt, starting the conversation with a greeting and no information on the code they wanted the LLM to produce. In each of these cases, what followed was a series of iterative attempts by the participant to incrementally clarify their intent, rather than rewriting the original prompt.  Several participants also struggled with understanding the problem and therefore wrote poorly formed prompts with little chance of achieving the desired result. In situations like these, many student conversations with ChatGPT were full of requests for ineffectual changes.

%
%
\floatstyle{ruled}
\newfloat{Listing}{htbp}{lop}
\begin{Listing}
\caption{Transcript of a conversation with ChatGPT.}
\label{listing:chat_1}
\small
\raggedright
\textbf{Prompt 1:} Hi, I want you to write a Python program.

\hfill \break
\textbf{ChatGPT response 1:} \textit{Sure, what would you like the program to do?}

\hfill \break
\textbf{Prompt 2:} I have a square matrix, and I want to swap the first half of the rows with the second half of the rows.

\hfill \break
\textbf{ChatGPT response 2:}\textit{ Here's a Python program that should do what you're asking for:}
\begin{minted}{python}
def swap_matrix_rows(matrix):
    n = len(matrix) half_n = n // 2
    for i in range(half_n):
        matrix[i], matrix[half_n + i] = 
        matrix[half_n + i], matrix[i] 
    return matrix
\end{minted}

\hfill \break
\textbf{Prompt 3:} That's not exactly what I meant. I actually want to swap the first half of the elements in each row with the corresponding elements in the second half of the rows.

\hfill \break
\textbf{ChatGPT response 3:} \textit{No problem, here's an updated version of the Python program that should swap the first half of the elements in each row with the corresponding elements in the second half of the rows:}
\begin{minted}{python}
def swap_matrix_elements(matrix): n = len(matrix)
    half_n = n // 2
    for i in range(half_n):
        for j in range(len(matrix[i]) // 2):
            matrix[i][j], matrix[half_n + i]
            [j - len(matrix[i]) // 2] = 
            matrix[half_n + i]
            [j - len(matrix[i]) // 2], 
            matrix[i][j] 
    return matrix
\end{minted}

\hfill \break
\textbf{Prompt 4:} That's not quite what I had in mind either. What I want to do is swap the first half of the columns in the first half of the rows with the last half of the columns in the last half of the rows.

\hfill \break
\textbf{ChatGPT response 4:} \textit{My apologies, here's an updated version of the Python program that should swap the first half of the columns in the first half of the rows with the last half of the columns in the last half of the rows:}
\begin{minted}{python}
def swap_matrix_elements(matrix): n = len(matrix)
    half_n = n // 2
    half_m = len(matrix[0]) // 2
    for i in range(half_n): for j in range(half_m):
        matrix[i][j], matrix[half_n + i]
        [half_m + j] = matrix[half_n + i]
        [half_m + j], matrix[i][j]
    return matrix
\end{minted}

\hfill \break
\textbf{Prompt 5:} Oops, sorry, I wasn't clear. I actually want to swap the first half of the elements in the first half of the rows with the corresponding elements in the last half of the rows.
\end{Listing}

Second, participants would prompt ChatGPT for code to solve the problem, paste the code into the automated assessment tool used in the exercise to check correctness, and then paste the failed test cases back into ChatGPT verbatim without any additional prompts or context. Participant \#12, after several prompts that resulted in code that was not quite complete, told ChatGPT: \emph{``This is very wrong, do the same as you had before but make sure that the top right quarter and the bottom left quarter are not edited by the function.''} ChatGPT responded with an apology and more almost-correct code. 

Third, some students misunderstood the assignment altogether, such as Participant \#13 who initially prompted ChatGPT: \emph{``give me a python 3 swap function which swap the the elements along the diagonal''.} After taking the code generated by ChatGPT and submitting it to the automated assessment tool used in the activity, they tried another incorrect prompt: \emph{``give me a python 3 swap function which works by first swapping the elements of each row in place, and then swapping the elements of each column in place.''} 

These examples indicate that many students, even ones many years into their programming education, do not necessarily understand how to write effective prompts.  They could benefit from explicit prompt writing practice that could teach them to understand the problem, write a single thorough prompt, and check the code generated by the LLM as having complete test case coverage.

\subsection{Reflections}
\label{sec:pilot-perceptions}

When reflecting on the task in our pilot study, many of the students mentioned that code producing LLMs need supervision to produce \textit{correct} code. After working with \mbox{ChatGPT} to produce correct code run against test cases, many students realized that writing code this way required a different kind of critical thinking. Instead of thinking through how to structure code to properly address the requirements, users instead need to carefully read and test the code generated for them to ensure it precisely fits their needs. 
Participants recognized that training was needed -- both training of novice students in how to use LLMs so that they are more of a help than a hindrance, and training of the models so that they provide responses targeted to novice learners.  For instance, participants said:


\begin{description}
\item P11: ``[It] made me reflect on the importance of critical thinking and proper supervision in using these models.''
\item P19: ``I think it's a double-edged sword in itself. Perhaps they need more supervision and training in the education field before they can become the best study companion for students outside of the classroom.''
\end{description}

Other students realized that prompt engineering to create code is a different kind of programming compared to how they learned. Coercing an LLM to generate correct code will need to be an iterative exercise in prompt refinement, not prompt conversation. Telling the LLM to revise its previous code, built from a previous prompt, may or may not work due to the way tools like ChatGPT maintain conversational context.
Examples of this from participants can be seen below:

\begin{description}
\item P12: ``I also found it interesting how difficult it was to get ChatGPT to write a somewhat simple program. I understand now that it is almost like a different kind of programming to get it to do what you want, and I can see the potential uses in education.''
\item P15: ``The most difficult part of this exercise is to properly instruct ChatGPT so that it could fully understand the requirements. ChatGPT appears to be `polite but stubborn', as it would generate code that could be run successfully but does not produce the correct output. When asking ChatGPT for a fix, it would politely provide a new snippet of code but the new code didn't effectively fix the problem, and sometimes there were no modifications made to the new code at all.''
\end{description}

These representative samples from the reflections by students indicated to us that learning how to successfully write prompts would need to be a skill taught explicitly in introductory programming courses, alongside other skills that are traditionally taught. We propose the idea of Prompt Problems to address this new gap in programming education.

%
%
\section{A tool for delivering Prompt Problems at scale: Promptly}

\begin{figure}
\centering
  \includegraphics[width=\linewidth]{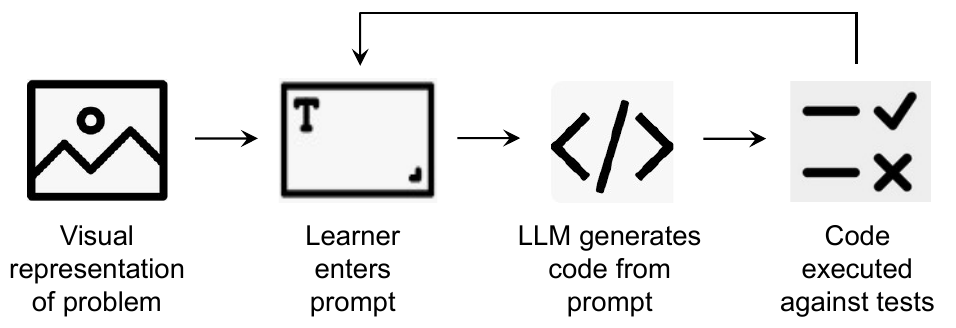}
  \caption{The \textsc{Promptly} tool implements a simple variation of Prompt Problems in which the code generated by the LLM is automatically executed against a set of test cases and can not be edited directly.  In order to modify the code, the learner is directed to edit the prompt.}
  \label{fig:schematic}
\end{figure}

We have developed a web-based tool called \textsc{Promptly} to support one particular variation of Prompt Problems, in which the code generated by the LLM is not editable by the learner (see Figure \ref{fig:schematic}).  Other variations of Prompt Problems are possible and we discuss these in Section \ref{discuss:variations}.

Our concrete implementation of the tool uses React and NodeJS as its key frameworks, and Material design for the styling of UI components. The client-side React implementation is accessible via Firebase Hosting, and the Express (NodeJS) backend is powered by Firebase Functions, operating within a serverless framework.  The backend communicates with OpenAI's API and transmits responses to a  JobeInABox\footnote{\href{https://github.com/trampgeek/jobeinabox}{github.com/trampgeek/jobeinabox}} sandbox which is hosted on an EC2 AWS instance. 
We explored the use of several specific OpenAI models, including \emph{text-davinci-003} and \emph{gpt-3.5-turbo}. Our current implementation uses \emph{text-davinci-003} which, although now officially a legacy model, is less likely to generate superfluous text and comments in the responses.  We found that the \emph{gpt-3.5-turbo} model requires significant additional prompting to increase the likelihood of generating only executable code, but that relying on prompting alone can be unreliable.  Future work will explore additional filtering approaches in order to transition to this newer model.
All relevant data, including prompts, responses and testing outcomes is stored using Firestore's NoSQL database.

%
%

\begin{figure*}
\centering
  \includegraphics[width=.8\linewidth]{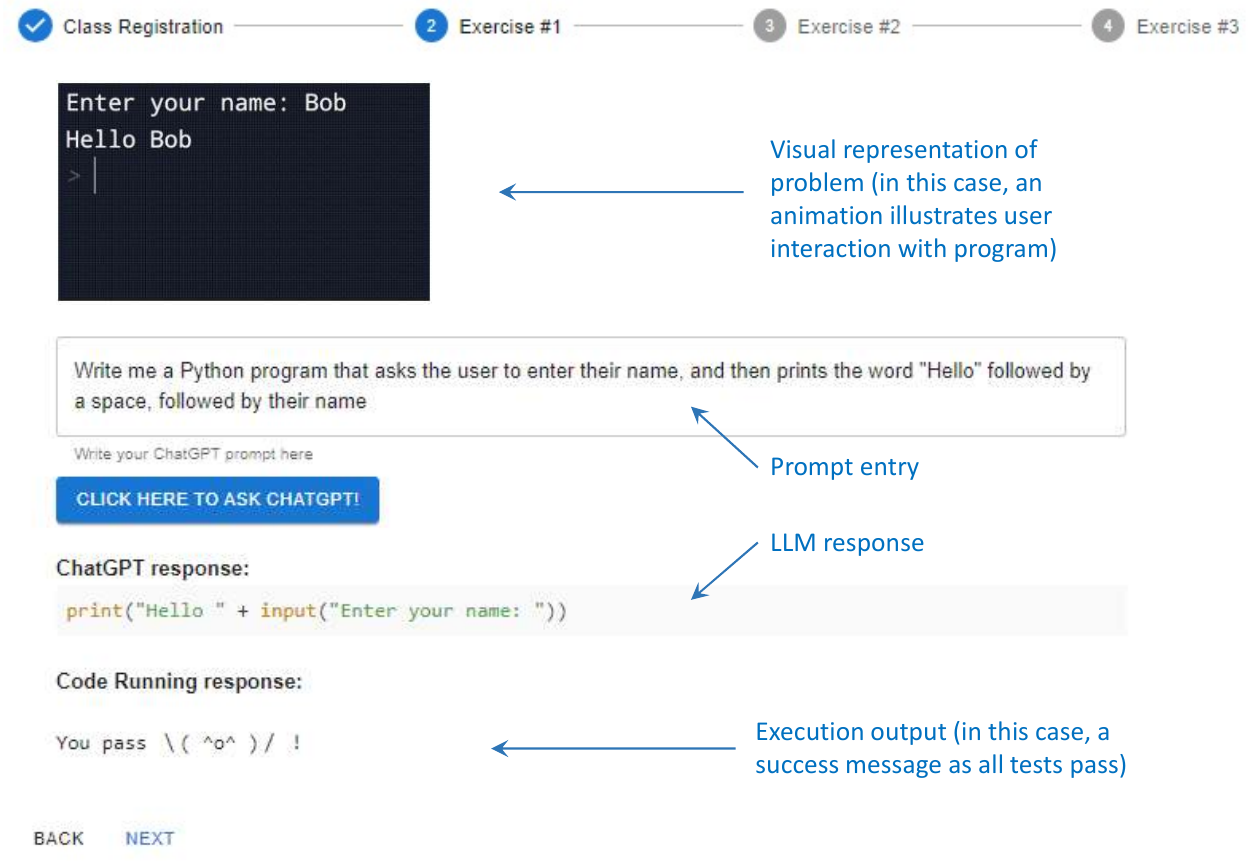}
  \caption{Interface layout for a Prompt Problem within the web-based \textsc{Promptly} tool (with figure annotations added in blue).}
  \label{fig:ex1_screenshot}
\end{figure*}

\subsection{Tool Design}

Within the \textsc{Promptly} tool, sets of Prompt Problems are organized into course repositories from which students can select after logging in.  Each Prompt Problem within a course repository consists of a visual representation of a problem -- that is, an image that does not include a textual description of the problem -- and a set of associated test cases that are used to verify the code that is generated by the LLM.  

Each set of Prompt Problems for a given course are presented in order, and a student can navigate through these using `Back' and `Next' buttons (see Figure \ref{fig:ex1_screenshot}).  Once a Prompt Problem is selected, the student is shown the visual representation of the problem, and a partial prompt to complete.  For problems where the solution is a Python program, this partial prompt begins: ``Write a Python program that...'', which provides guidance to the student.  If the problem requires students to write a single function, then the partial prompt is: ``Write a Python function called...''.  As soon as any text for extending the prompt is entered by the student, the ``Click here to ask ChatGPT!'' button is enabled.  Clicking this button constructs a prompt that is sent to the LLM. This prompt consists of the verbatim text entered by the student, as well as some additional prompting to guide the model to produce only code and no additional explanatory text.  

Once the code response is received from the LLM, it is then sent to a sandbox for execution against the set of test cases.  We use the publicly available sandbox associated with the CodeRunner tool \cite{lobb2016Coderunner}.  If the generated code passes all of the test cases for the prompt problem, then the student receives a success message and is directed to progress to the next problem.  If any of the test cases fail, then the first failing test case is shown to the student.  At this point, they are able to edit the prompt and resubmit in order to generate a new code response. 

Figure \ref{fig:ex1_screenshot} shows a screenshot of the tool interface once the learner has logged in and selected their course.  The following instructional message is shown but not included in the screenshot: \emph{``Your task is to view the visual representation of the problem and then type a prompt which describes the task sufficiently well for the language model to generate a correct solution in Python. If the code that is generated is not correct, you will see test output below the coding area and you can try again by modifying the prompt!''}.  In the screenshot in Figure \ref{fig:ex1_screenshot}, the first problem in a sequence of three problems for the course is shown, and has just been solved by the learner.

\begin{figure}
\centering
  \includegraphics[width=\linewidth]{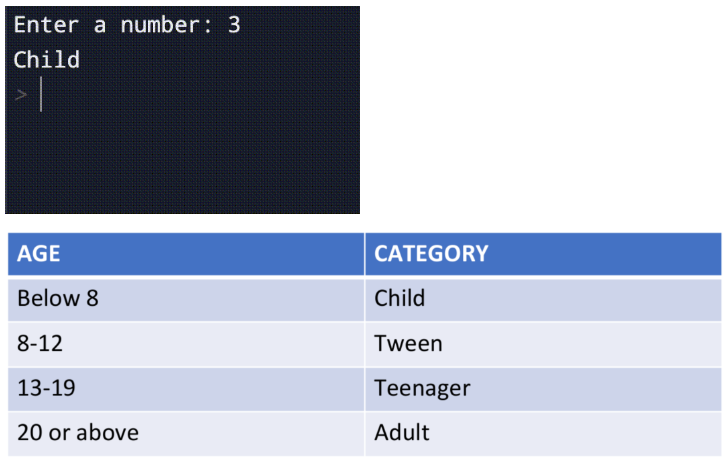}
  \caption{Producing a categorization based on age.}
  \label{fig:ex2_question}
\end{figure}

\begin{figure}
\centering
  \includegraphics[width=\linewidth]{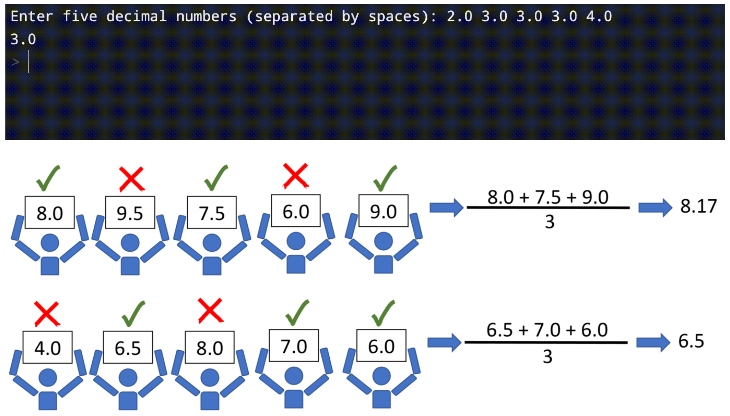}
  \caption{Calculating the average of the ``middle'' values out of a set of five values (using the metaphor of judges scoring an athletic competition, where the highest and lowest values are excluded).}
  \label{fig:ex3_question}
\end{figure}



\subsection{Classroom Evaluation}
%
%
Prompt Problems are a novel type of activity for learners in an introductory programming course, and so we are interested in understanding their perceptions of the \textsc{Promptly} tool, and on their interactions with it when solving problems.
We organise our investigation of the way students use \textsc{Promptly} around the following two research questions:

\begin{description}
\item RQ1: How do students interact with the \textsc{Promptly} tool in terms of overall success rates and on the lengths of the prompts they construct?
\item RQ2: What are students' perceptions of the \textsc{Promptly} tool and on learning programming through constructing prompts for LLMs?
\end{description}


To explore these questions, we deployed \textsc{Promptly} as an ungraded (i.e. optional) laboratory task in a large introductory Python programming course taught at the University of Auckland.  Students in this course typically have no prior programming experience.  The lab was conducted in the second week of the course, at which point students were writing single-file scripts, without the use of functions, and had learned about standard input and output, arithmetic, and conditional statements.  


Three problems were available on \textsc{Promptly} for students to attempt.  The first of these was the problem previously illustrated in Figure \ref{fig:ex1_screenshot}, where the goal was to write a program that would ask the user to enter their name, accept input from the command line, and then display ``Hello '' followed by the name as standard output. 
The other two problems are illustrated in Figures \ref{fig:ex2_question} and \ref{fig:ex3_question}.  The second problem (Figure \ref{fig:ex2_question}) required a program that accepts an integer input from the user representing an age, and then prints a textual cateogrization of that age.  The third problem (Figure \ref{fig:ex3_question}) required a program that accepted five floating point inputs and then calculated the average of the three middle values (i.e. after removing the maximum and minimum values). 

For all three problems, the visual representation of the problem included a short animated image ($\sim$10 second duration), shown as a command-prompt style window.  The command-prompt animation illustrated entry of user input, one key at a time, and then the subsequent display of output from the program.  For the second and third problems, the visual representation also included a corresponding image that highlighted several pairs of inputs with their corresponding output. 

In terms of interactions with the tool (RQ1) we calculate, for each of the three problems, the average number of prompt submissions that were required to solve it, the number of students who were successful, and the average number of words used in the submitted prompts. 
To investigate student perceptions of the activity and tool (RQ2), students were invited to provide feedback on their experience using \textsc{Promptly}.  This feedback was not graded, and was given in response to the following prompt: \emph{``We would appreciate hearing about your experiences completing the exercises and in particular, how you think the experience of writing prompts may help you to learn programming''}.


\section{Results}

Our study was conducted in July 2023, and participation by students was optional.  A total of 54 students attempted at least one problem on \textsc{Promptly}, which represents approximately 12\% of the enrolled students in the course.  

\subsection{Student interactions with \textsc{Promptly}}

We measured several performance indicators around student use of \textsc{Promptly}, such as prompt lengths and number of submissions.  As summarized in Table \ref{tab:summary}, on average participants submitted 2.70 attempts for problem 1, 2.16 submissions for problem 2, and 6.4 submissions for problem 3.   On this basis, problem 3 appeared to be the most difficult for students, and this is further supported by student reflections (which are reported in more detail in Section \ref{sec:reflections}), with one student commenting: \emph{``The instruction for the third exercise is not clear I don't get it.''}  Listing \ref{listing:prompts_students} shows three prompts that were submitted by different students attempting problem 3.  Some students found it difficult to infer the goal from the problem representation.  For example, the first prompt shown in Listing \ref{listing:prompts_students} is an example where the student has incorrectly inferred that values included in the average calculation should be sufficiently close to their predecessors.  Trying to generate code for the wrong problem can be frustrating, which may explain the following part of the prompt written by the student: \emph{``If the user has not provided numbers that sufficiently meet this criteria, call them an idiot''}.

In the second example in Listing \ref{listing:prompts_students}, the student has not attempted to provide a prompt that demonstrates they have understood what the problem is asking, but instead they have created a prompt that simply parrots back to the tool the three example tests cases shown in the problem description.  The student then asks the model: \emph{``Can you please replicate this program?''}.  The student submitted this prompt four times in a row, but all attempts were unsuccessful.
Finally, the third example in Listing \ref{listing:prompts_students} is the shortest successful prompt that was submitted for this problem.  

Overall, the average number of words in prompts for each of the problems was 13, 38, and 36, respectively. The number of students that solved the problems was 43, 32, and 19, respectively. 

\begin{figure}
\centering
  \includegraphics[width=\linewidth]{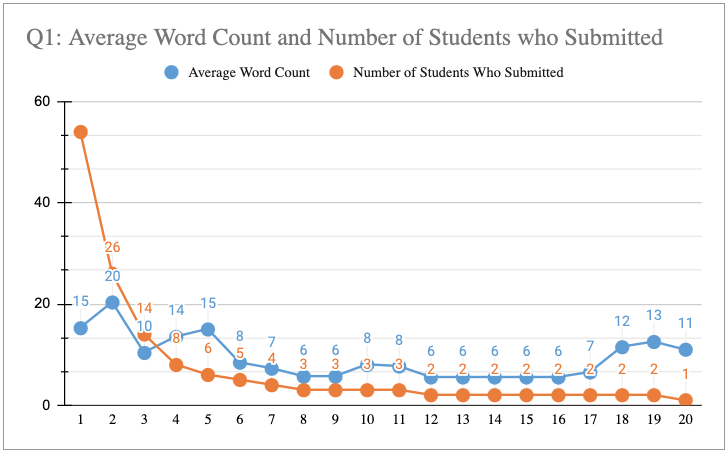}
  \caption{The average number of words in each subsequent submission for problem 1 compared to the number of participants that submitted.}
  \label{fig:q1_avgwords-vs-numStudentSub}
\end{figure}

\begin{figure}
\centering
  \includegraphics[width=\linewidth]{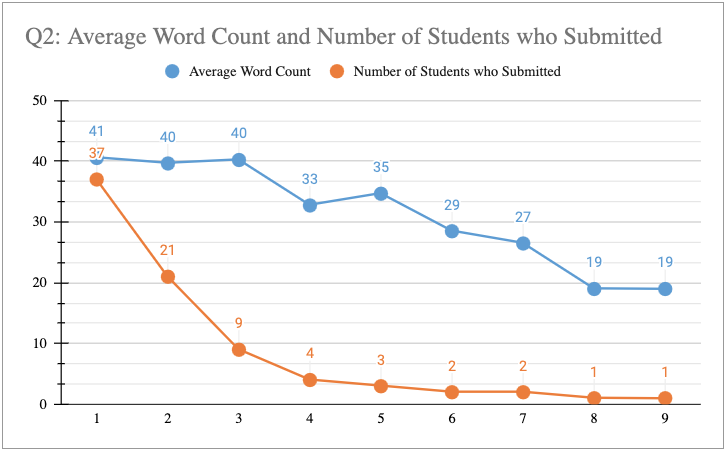}
  \caption{The average number of words in each subsequent submission for problem 2 compared to the number of participants that submitted.}
  \label{fig:q2_avgwords-vs-numStudentSub}
\end{figure}

\begin{figure}
\centering
  \includegraphics[width=\linewidth]{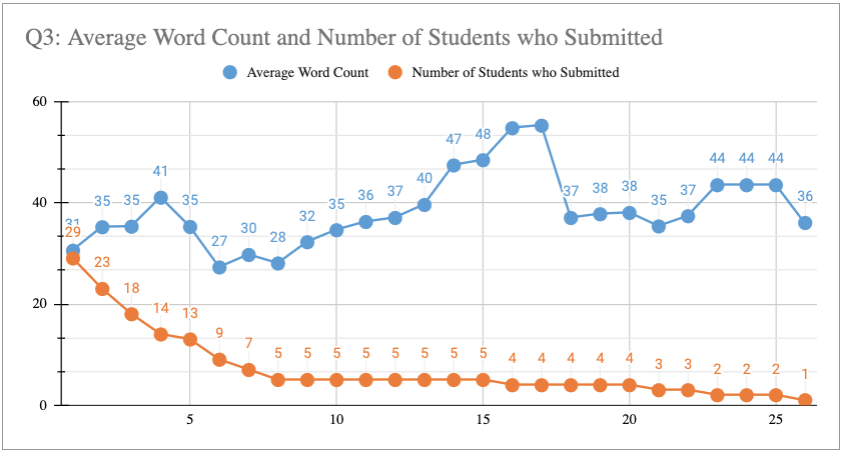}
  \caption{The average number of words in each subsequent submission for problem 3 compared to the number of participants that submitted.}
  \label{fig:q3_avgwords-vs-numStudentSub}
\end{figure}

Figures \ref{fig:q1_avgwords-vs-numStudentSub}, \ref{fig:q2_avgwords-vs-numStudentSub} and \ref{fig:q3_avgwords-vs-numStudentSub} illustrate, for each of the three problems, trends regarding how the average word count of prompts, and the number of students writing them, change across subsequent submissions.   For example, the left most column in Figure \ref{fig:q1_avgwords-vs-numStudentSub} shows that 54 students made an initial submission to this task and that on average, the word length of all of these submitted prompts was 15.  As students either solve or abandon the problem, fewer students make subsequent submissions.  
Comparing the three problems, prompt lengths tend to decrease over subsequent submissions for problems 1 and 2, but tend to slightly increase for problem 3.

%
%
\floatstyle{ruled}
\newfloat{Listing}{htbp}{lop}
\begin{Listing}
\caption{Three student-submitted prompts for Problem 3}
\label{listing:prompts_students}
\small
\raggedright
\textbf{Misinterpreting the problem:} \break
Write me a Python program that does the following:\break
1. Prompts the user to enter five decimal numbers (1dp) between 1.0 and 10.0 separated by spaces. \break
2. Chooses three of these numbers using the following rules: a number chosen be different from the previously chosen numbers and each subsequently chosen value must be within 0.5 of its predecessor. If the user has not provided numbers that sufficiently meet this criteria, call them an idiot and prompt them for another five values. \break
3. Find the average of these numbers and round the result to 2dp. Precede this result with the numbers chosen.

\hfill \break
\textbf{Parroting the tests:} \break
A Python program requests the user "enter five decimal numbers (separated by spaces)".  In the first example the user inputs the five numbers 2.0 3.0 3.0 3.0 4.0 to which the program outputs 3.0.  In the second example the user inputs the five numbers 8.0 9.5 7.5 6.0 9.0 to which the program outputs 8.17 .  In the third example the user inputs the five numbers 4.0 6.5 8.0 7.0 6.0 to which the program outputs 6.5. Can you please replicate this program?

\hfill \break
\textbf{Successful:} \break
Write me a Python program that takes five decimal number separated by spaces, and outputs the average of the 3 median numbers as a decimal rounded to 2dp.

\end{Listing}



\begin{table}[htb]
\caption{Summary of usage of \textsc{Promptly}.  For each question, the average number of submissions required to solve the problem is shown, along with the number of students who successfully solved the problem and the average number of words in prompts submitted.}
\begin{tabular}{cccccc}

{\bf Problem} &	{\bf Average} & {\bf Students} & {\bf Average} \\
{\bf id} &	{\bf submissions} & {\bf solved} & {\bf words} \\
\toprule
1 & 2.7 & 43 & 13  \\
2 & 2.2 & 32 & 38  \\
3 & 6.4 & 19 & 36  \\
\bottomrule
\label{tab:summary}
\end{tabular}
\end{table}

\subsection{Student reflections on \textsc{Promptly}}
\label{sec:reflections}
We analyzed feedback from 58 students who provided a response to the reflection question.  This is slightly greater than the number of students who used \textsc{Promptly}, but as we report below, some students indicated a resistance to using LLMs for generating code and thus responded to the reflection statement without using the tool.  We report the main themes that emerged from our analysis below. 

\subsubsection{Exposure to new coding constructs}

Given that our evaluation was conducted early in the course, the code that was generated would sometimes contain features that were unfamiliar to students.  For the most part, students commented positively on this aspect, and a theme emerged around the way the tool introduced students to new programming constructs and techniques.  As one student commented: \emph{``These exercises introduced me to new functions... so this method of writing code could help increase my programming vocabulary''}.  Another response aligning with this theme was: \emph{``Honestly that was pretty cool, I like the way it works and how we can see the code afterwards. Even if we don't know how to code some of the features, knowing the steps and then seeing how it's meant to be done is very helpful''}.

One student commented on the value of seeing both the structure and syntax of the code generated by the LLM: \emph{``The main benefit I gained from using ChatGPT in this environment was observing the logical structure of the programs that it created to fulfill the established requirements. In all three cases it used functions that I was previously unaware of, allowing me to gain an understanding of how they could be used and the correct syntax for implementing them.''}

\subsubsection{Enhancement of computational thinking}

We also found students valued the way in which the tool challenged them to think carefully about how to solve a problem and communicate precise specifications: \emph{``You would have to understand what the question is asking to be able to give a working prompt to the AI, so it seems very good for making you think properly about the question''}.  Writing clear prompts can involve communicating problem solving steps, and thus draws on computational thinking skills.  This is illustrated well by the quote: \emph{``Writing prompts helps learn programming by making you have to think about what the task at hand and to split it up into segments of which you need to describe to the AI... I would say that this would help students with the act of breaking down a big project into little tasks.''}.  Another similar response highlighted how the tool helped in visualizing the problem solving process:  \emph{``Writing the prompts can help you with visualizing the steps required in the programming''}.  

\subsubsection{General Positive Feedback}

Around one-third of the participants expressed generally positive sentiments about the \textsc{Promptly} tool, and this was the most common theme overall. Representative comments include: \emph{``I think that is was a good for practicing asking AI''} and \emph{``Asking AI to write promps help a lot in programming!!!''}.  One student who viewed the tool positively, also expressed some concern about the power of the underlying code-generating models: \emph{``It is absolutely a great tool, however in this regard it's kind of terrifying. It was able to process my instructions fluently.''}

Some students also commented more generally about the importance of learning how to use generative AI responsibly in computing courses, and the value of having explicit approaches for teaching this.  For example, \emph{``I think it is very smart ... to work on a way to integrate and teach a responsible usage of machine learning!''} and \emph{``I think it would be very useful to have a section of a course that teaches how to use prompts properly and provide a better understanding of AI tools.''}

\subsubsection{Resistance and negative feedback}

Although less common, we found that some students appeared resistant to using the tool, citing fears about potential impacts on their creativity.  One student expressed: \emph{``I don't have much intention of using ChatGPT at the moment as I major in design and I have a strong belief in personal creativity''}.  Another was more blunt: \emph{``I refuse to use chatGPT for programming''}.  Over-reliance on AI generated outputs is a commonly cited concern within the education community, and several students commented on this aspect, including: \emph{``it is critical for students to learn the ability to write code independently rather than relying only on AI-generated answers''} and \emph{``I feel like it is too tempting of a tool to use through the labs and not learn and develop these skills yourself''}.  Further exploring these concerns would be an interesting avenue for future work.

Overall, while most students reported finding \textsc{Promptly} beneficial, particularly for exposure to new programming constructs and for strengthening computational thinking skills when communicating a problem, a minority of students were hesitant about the use of generative AI tools for learning programming.

%
%
\section{Discussion}
%
%




Beginning typically with very small problems in CS1 and proceeding on to semester- or year-long applied problems in software engineering, CS curricula expose students to both top-down and bottom-up approaches. 
The problems presented in \textsc{Promptly}
can be considered to be ``bottom-up'', since students start with input-output pairs and have to infer a problem description.  And yet, the prompts that the students write can be considered ``top-down'' since the system requires students to abstract the meaning of the input-output pairs into English sentences and not code.  Students need to understand the problem before they can correctly generate prompts that cause the LLM to produce correct code.

In contrast to other tools students use, such as compilers, learning to use LLMs presents unique challenges. 
Although the literature continues to document the difficulty students have with compiler error messages, one thing we have never worried about teaching students is that compilers might sometimes just get it wrong. In contrast, at this point in time, LLMs sometimes generate answers that are syntactically and semantically incorrect.  
Deliberate exposure to the inconsistencies of outputs generated by LLMs can serve to highlight the importance of a ``critical eye'' in evaluating generated code and may help to moderate the potential for over-reliance on these tools.
The use of LLMs to generate code from prompts places the responsibility for ensuring correctness on the user, so adequate testing becomes more important.
Future tools that focus on prompt generation would benefit from the integration of user-generated tests to ensure students are explicit about the program requirements when they create prompts. Tasking students with generating test cases (before writing code) has previously been studied as an approach to help improve problem understanding \cite{denny2019closer, pechorina2023metacodenition}.

It is worth noting that our tool does not provide instruction for students about how to create prompts effectively.  It merely requires them to complete a task that involves prompt creation.  This is aligned with most other Automated Assessment Tools that provide assessment tasks to support learning, but little explicit instruction~\cite{keuning2018systematic}.  Neither the students in our pilot study nor those in our classroom evaluation of the tool were taught prompt creation, so currently we have focused on students' intuitions around prompt generation.  Future work will explore how to more directly teach students to generate prompts in structured ways to determine if such instruction positively impacts their performance using tools that assess prompt generation.

Although the current system evaluates prompt effectiveness in producing correct programs, it does not evaluate the efficiency of the prompts. Our unit tests consider only whether the given inputs are translated to the expected outputs.  A prompt could include irrelevant words and generate irrelevant code constructs, and as long as it still translates the given inputs to the expected outputs, our system will treat the task as completed successfully.  Future work must address how to go beyond effective prompts to efficient (and effective) prompts.


\subsection{Variations}
\label{discuss:variations}

Prompt Problems are a class of problems where learners generate prompts that are given to LLMs to produce code.  There are various ways that such problems can be implemented, and several considerations for designing them.  Our tool currently makes certain implementation trade-offs.  It does not allow dialogue, it does not allow students to edit the code that is generated by the LLM, and it evaluates only a single solution at a time. 

\subsubsection{No dialogue}
ChatGPT interfaces include support for ``dialogue''.  This interaction style is natural and easy on the beginner.  The reason we did not support this interaction and forced the student to ``start from scratch'' each time is that we were deliberately focusing the student on creating a complete, top-down, problem description.  Although it is more cognitively demanding to require the student to provide all relevant information in a single prompt, we were interested in teaching exactly that process.

\subsubsection{No access to the code}
Although students who are more advanced may find it easier to simply write the code than construct a prompt, our interest is in providing students with experience in generating prompts.  For this reason, we did not allow students to edit the code that was generated.  We did show the code and students were able to study the generated code along with the unit test results to modify their prompts for another submission.  Our system is not intended to be a realistic IDE for code development, but future tools that support Prompt Problems could allow code editing to provide refactoring practice and a more authentic experience.  

\subsubsection{Single solution generated}
LLMs generate different variations of output for the same input prompt.  Our tool does not currently address the possible variation in generated content.   Prompts can be brittle, as sometimes the same prompt may work and other times it may not.  
Non-deterministic behaviour of the models may be frustrating for students, as simply resubmitting a previously unsuccessful prompt may actually work.  Nevertheless, this may be a useful learning experience for students, as it helps to highlight this inherent LLM behaviour.  A different variation of the current tool design could generate multiple code implementations every time a single prompt is submitted, allowing students to compare them and see which ones satisfy the problem.  Viewing multiple correct, but different, implementations of the same algorithm is useful for helping students understand that there are frequently many correct solutions to a problem \cite{luxton2013differences}. Future work could explore how to present this aspect of LLMs to students who are learning to write prompts.


 \begin{figure}
\centering
  \includegraphics[width=\linewidth]{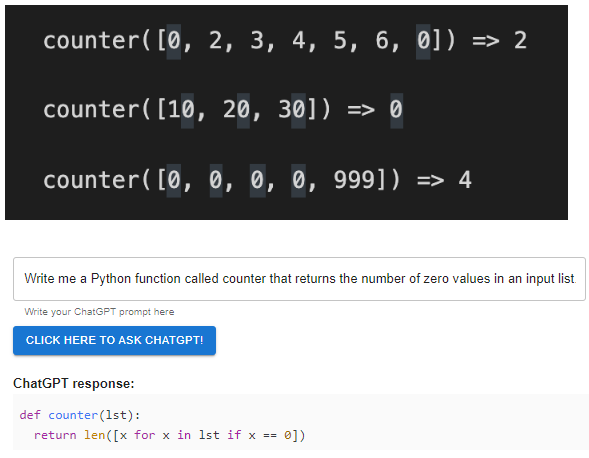}
  \caption{A problem in \textsc{Promptly} represented as a set of input-output pairs, where the solution requires generating a function (called `counter') that returns a count of the number of times zero occurs in a list. One possible prompt, and the resulting code that correctly solves the problem, is shown. }
  \label{fig:function_example}
\end{figure}

\subsection{Problem Design}
Our restriction on ``visual'' problem representation is motivated by a desire to prevent students from simply copying-and-pasting into the LLM prompt.  
The most important feature of the problem representation is that it does not provide the text that can be directly used as a prompt.  For problems where the desired code output is a function, listing a set of test cases (input and output pairs) is a convenient representation.  In our current implementation of \textsc{Promptly}, this can be achieved with plain text, or with an image to discourage copy-and-paste behaviour, as illustrated in Figure \ref{fig:function_example}.  There are several considerations for how to design a Prompt Problem which we discuss here.


\subsubsection{Avoiding textual clues}

One of the biggest limitations in terms of preparing Prompt Problems is that as an instructor, you have to think about tasks that can be presented visually to learners.  Even simple tasks such as ``Order the elements of the list in ascending alphabetical order'' which can be explained in few words, is quite challenging to convey visually without the use of text. Without a textual explanation, students are expected to use \textit{inductive} reasoning to determine what problem is being solved from visual examples that typically include specific cases.  As we found, this can be quite challenging in some cases.  For example, many students found problem 3 challenging in our classroom evaluation, with one commenting on their reflection: \emph{``The last question seemed unnecessarily unclear, I can't imagine there will be many instances where the task I'm meant to do will be as vague as what was given in question 3''}.  In this evaluation, we used short animations to illustrate data being entered at the command prompt.  Although such animations can convey additional information, they are more difficult to create. 

\subsubsection{Accessibility}
Educational resources should be accessible to students with a visual impairment.  This is typically satisfied with a text-based description of visual media which can be read aloud.  However, if a text-based description of the image is provided, then this may either (a) be sufficiently descriptive of the problem that it could be passed directly to an LLM without requiring a student to engage with the prompt construction strategy; or (b) add a further layer of complexity to the inductive reasoning required to determine the problem that is being illustrated by the visualization.  For example, Figure~\ref{fig:ex3_question} is intended to convey that a program should accept 5 numbers and remove the highest and lowest values before calculating the average of the central 3 values.  However, a textual description of the image may focus undue attention on the many details that provide context, but which are not directly related to the problem.

\subsubsection{Natural language bias}
Students for whom English is their native language may, in general, be able to produce prompts in English that are more nuanced in their use of language, and are likely to have greater success in improving partially correct prompts.  Students with more limited English language could be disadvantaged in manipulating the LLM to produce the correct program, even when they understand the problem and the programming solution more effectively than a native English speaker.  Instructors who plan to use prompt generation activities as part of formal graded assessment should consider the extent to which English language skills should impact grades in their course.

\subsubsection{Prompts and specificity}
Creating a prompt that gives a general description of the problem is reasonably straightforward, but as instructors are aware, being precise and complete when describing the requirements for a problem relies on experience and expertise.  Students are typically very familiar with following the specifications of a problem, but are often less familiar with the process of specifying desired functionality with precision. For example, our pilot study (see Section~\ref{sec:pilot_study}) revealed that graduate students were frequently not providing sufficient information in their prompt to the model.  
Similarly, traditional code writing exercises do not encourage students to think about corner cases, because these are typically provided in the problem description (usually carefully worded by an instructor) or shown in test case output.
This suggests that explicitly training prompt construction, as we propose, may make a valuable contribution to computing education by focusing more attention on important dispositions, such as being precise and paying attention to detail.




\subsubsection{Inappropriate solutions}
When solving Prompt Problems, the LLM might produce code which is too advanced relative to the timing of the course, and we may not wish to show this to learners.  This could be both  negative and positive --- it might show students new approaches they have not seen before, but on the other hand it could be confusing and demotivating as students may feel like they should understand the code when they do not.  For example, in our classroom evaluation, although most students commented positively on this aspect, we did see some evidence of students being confused by the outputs: \emph{``when the question prompt got harder, the code become harder as well and I wasn't able to understand the code that was being generated''}, and \emph{``some of the functions used in the latter exercises were new to me and I would not be able to diagnose any code errors within it''}.  One way of handling this issue could be through tool design, by including in the tool filters for certain programming constructs that should be used for given problems (instructors could define these along with the problems).  These filters could either be post-filters (i.e. rejecting a model completion and requesting a new one if it includes concepts that are not desired) or pre-filters (i.e. where the prompt is modified to include which constructs are allowed).



\subsubsection{Problem difficulty}
Prompt creation is a new kind of task that we (as a community) have limited experience with, and we have not typically asked students to complete similar tasks.  It may be difficult for instructors to have an intuition for how hard it will be for students to construct prompts for various problems.  In addition, further thought is needed about when to introduce such tasks into the curriculum.  Novices in a typical CS1 course could potentially solve more complex problems earlier than they would otherwise if they had to generate code from scratch.  However, it may be useful for students to have some minimal knowledge of programming in order to be able to diagnose problems in code generated by LLMs.

\section{Conclusion}

In this work we present a novel pedagogical approach, known as `Prompt Problems', designed to help students learn how to craft effective prompts for generating code using large language models (LLMs).  This is an essential skill in the current era of rapidly advancing AI and automated code generation.  Learning effective prompt construction is important as it can help students express detailed specifications, encourage them to think about corner cases and apply computational thinking skills. Indeed, we motivate our work by presenting the findings from a pilot study involving graduate students which revealed struggles in providing sufficient details when writing prompts.


We make three primary contributions in this paper.  The first is the conceptualization of Prompt Problems as a nascent pedagogical strategy.  The second is the design and implementation of a novel tool, \textsc{Promptly}, for delivering Prompt Problems at scale.  The third contribution is an empirical evaluation of \textsc{Promptly} in a first-year Python programming course, where we explore student interactions with and perceptions of the tool.  
Future research should investigate different variations of the approach we have described, including permitting code-editing and dialogue-based interactions, which present both benefits and challenges.  It is also essential to explore the right time to introduce students to the concept of prompt-based code generation, and how to integrate these problems in parallel with conventional teaching practices. 


\bibliographystyle{ACM-Reference-Format}
\bibliography{main.bib}

\end{document}